\documentclass[prl,twocolumn,showpacs,amsmath,amssymb,superscriptaddress]{revtex4}

\usepackage{amsmath}
\usepackage{amsfonts}
\usepackage{amssymb}
\usepackage{color}
\usepackage{units}
\usepackage{verbatim}
\usepackage{graphicx}
\usepackage[normalem]{ulem} 
\usepackage{placeins}

\newcommand{\beq}{\begin{equation}}
\newcommand{\eeq}{\end{equation}}
\newcommand{\bqa}{\begin{eqnarray}}
\newcommand{\eqa}{\end{eqnarray}}

\newcommand{\blk}{\color{black}}

\definecolor{ngreen}{rgb}{0.2,0.6,0.2}

\definecolor{golden}{rgb}{0.8,0.6,0.1}

\definecolor{purp}{rgb}{0.8,0.1,0.8}

\definecolor{orange}{rgb}{0.9,0.3,0}
\definecolor{mar}{rgb}{0.6,0.1,0.1}

\renewcommand{\section}[1]{{\em #1}.---}

\begin{document}

\title{Heralded Noiseless Amplification of a Photon Polarization Qubit}

\author{S. Kocsis}
\affiliation{Centre for Quantum Computation and Communication Technology (Australian Research Council)}
\affiliation{Centre for Quantum Dynamics, Griffith University, Brisbane, 4111, Australia}
\author{G. Y. Xiang}
\affiliation{Centre for Quantum Dynamics, Griffith University, Brisbane, 4111, Australia}
\affiliation{Key Laboratory of Quantum Information, University of Science and Technology of China, CAS, Hefei 230026, China}
\author{T. C. Ralph}
\affiliation{Centre for Quantum Computation and Communication Technology (Australian Research Council)}
\affiliation{Department of Physics, University of Queensland, Brisbane 4072, Australia}
\author{G. J. Pryde} \email{G.Pryde@griffith.edu.au}
\affiliation{Centre for Quantum Computation and Communication Technology (Australian Research Council)}
\affiliation{Centre for Quantum Dynamics, Griffith University, Brisbane, 4111, Australia}

\begin{abstract}

Non-deterministic noiseless amplification of a single mode  \cite{Xiang10,Ferr10,Zav10,Usuga10,Osorio12} can circumvent the unique challenges to amplifying a quantum signal, such as the no-cloning theorem \cite{WootZur82}, and the minimum noise cost for deterministic quantum state amplification \cite{Caves81}. However, existing devices are not suitable for amplifying the fundamental optical quantum information carrier, a qubit coherently encoded across two optical modes. Here, we construct a coherent two-mode amplifier, to demonstrate the first heralded noiseless linear amplification of a qubit encoded in the polarization state of a single photon. In doing so, we increase the transmission fidelity of a realistic qubit channel by up to a factor of five. Qubit amplifiers promise to extend the range of secure quantum communication \cite{Gisin10,Minar12} and other quantum information science and technology protocols.

\end{abstract}

\maketitle

Photons are the best long-range carriers of quantum information, but the unavoidable absorption and scattering in a transmission channel places a serious limitation on viable communication distances. Signal amplification will therefore be an essential feature of quantum technologies, with direct applications to quantum communication, metrology, and fundamental tests of quantum theory. The quintessential model for encoding quantum information is the qubit. Qubits, or systems of entangled qubits, are central to most protocols for transmitting and processing quantum information \cite{NielsenChuang}, and play a large role in other proposed quantum technologies \cite{Shaji08,Gao10} and proposed investigations of quantum mechanics (\emph{e.g.} \cite{Kwiat94}). A natural implementation of a travelling qubit is one excitation shared between two harmonic oscillators. (This implementation may also be relevant to cavities or other bounded oscillators). In optics, this implementation is a \emph{photonic qubit}, in which the information is encoded in orthogonal polarization, spatial or temporal modes of a single photon.

\begin{figure}[!h]
\includegraphics[width=\columnwidth]{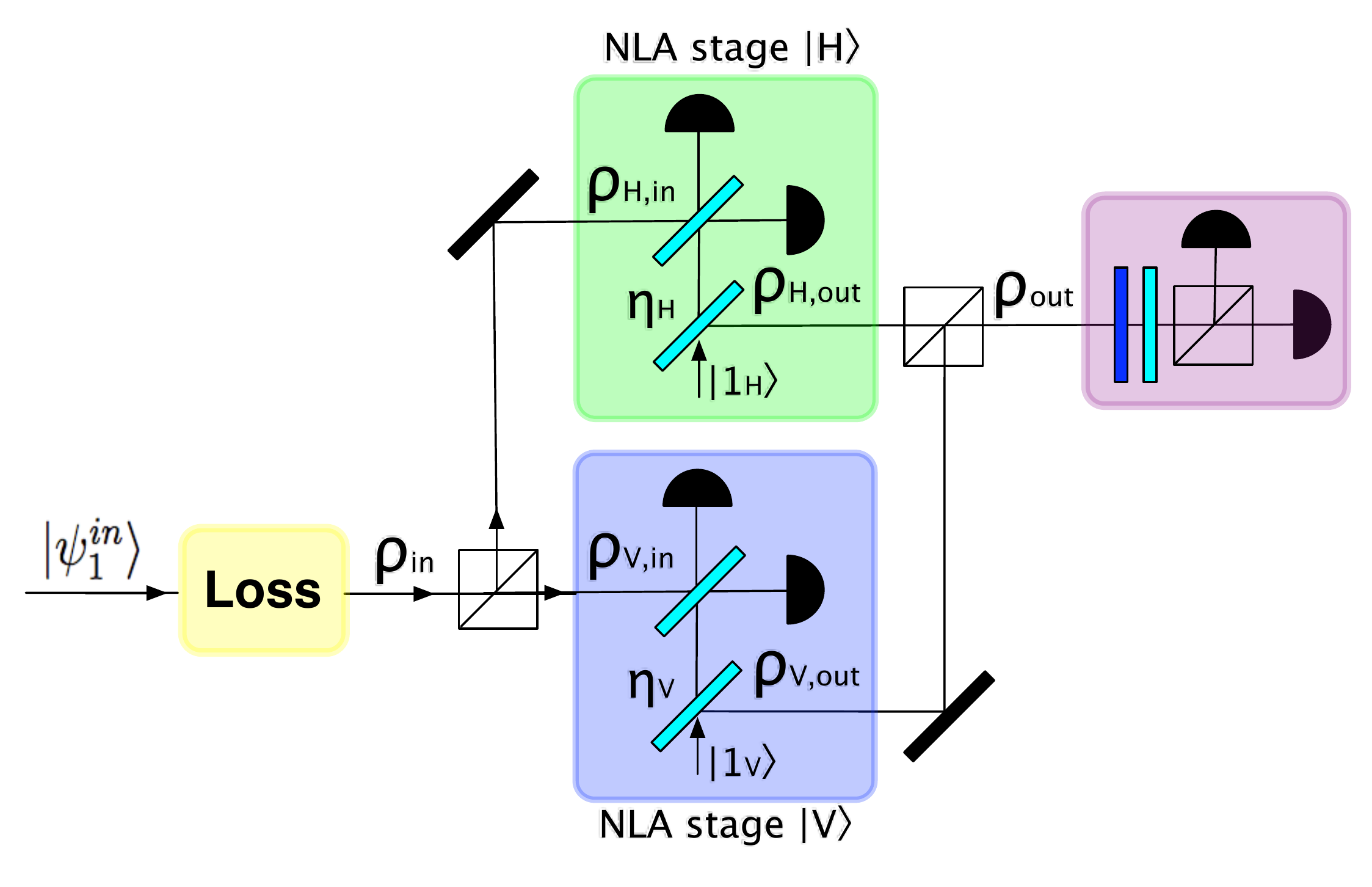}
\includegraphics[width=\columnwidth]{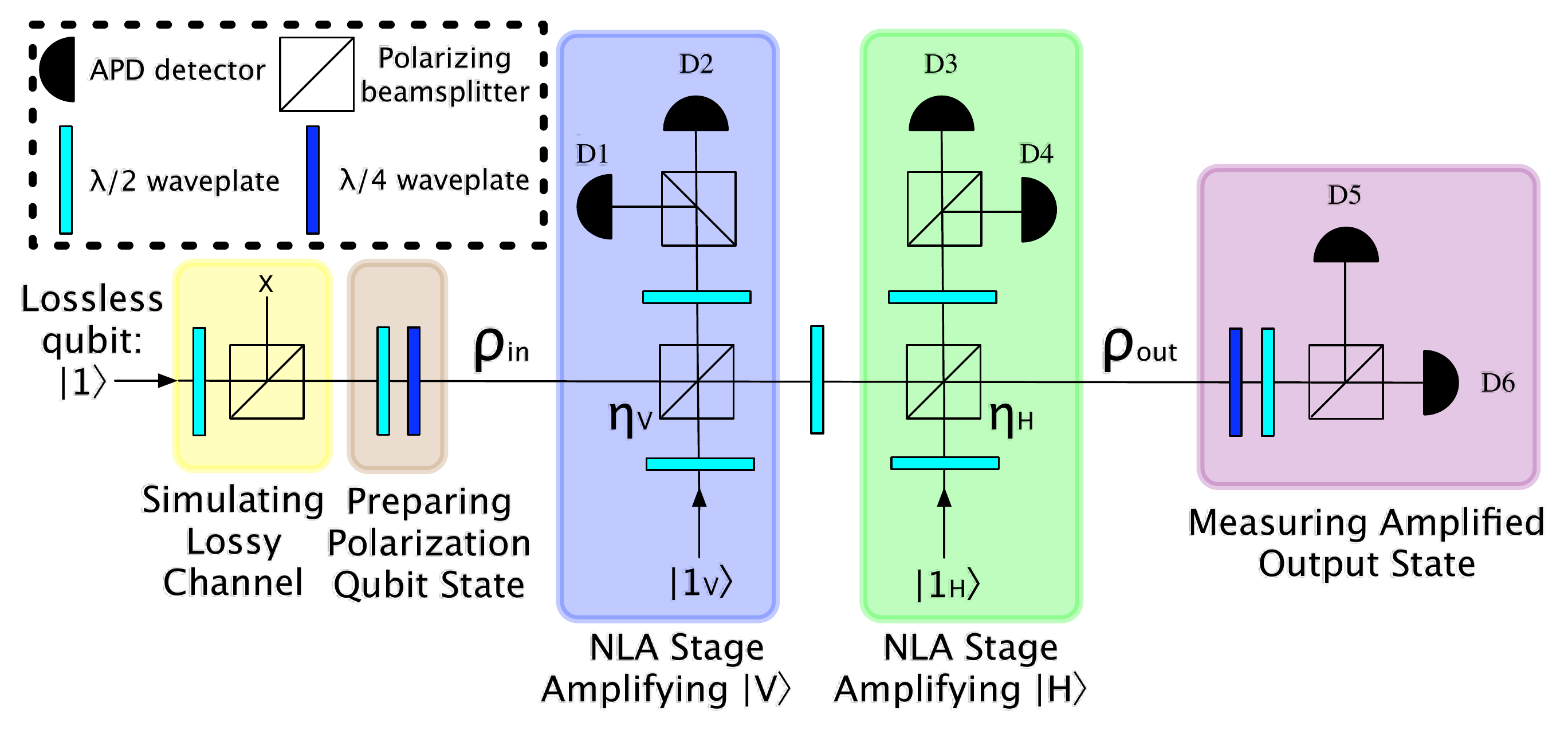}
\caption{{\bf{Top:}} Conceptual representation of a qubit amplifier circuit. The input signal $\rho_{in}$ is split with a polarising beamsplitter into its polarisation components $\rho_{in,H}$ and $\rho_{in,V}$, which are individually amplified by separate NLA stages. These NLA stages work by the generalised quantum scissors \cite{Xiang10}, as shown in the coloured boxes. The reflectivities $\eta_{H}$ and $\eta_{V}$ (always set to be equal) of the variable reflectivity beamsplitters are related to the amplifier gain through $g^2 = \eta/(1-\eta)$. The outputs from the two NLA stages, $\rho_{out,H}$ and $\rho_{out,V}$ are coherently combined to recover $\rho_{out}$, the amplified qubit. {\bf{Bottom:}} The experimental realisation of the qubit amplifier. The variable reflectivity beamsplitters are implemented with half wave plates and polarising beamsplitters. Here, the interferometer from the top figure is achieved in polarisation; $|V\rangle$ and $|H\rangle$ are amplified by two NLA stages in series, and recombined at the output in a way that is inherently stable. We implemented the loss prior to qubit state preparation. This is identical to polarization-independent loss after qubit preparation.}
\end{figure}

\begin{figure*}[t]
\includegraphics[width=\textwidth]{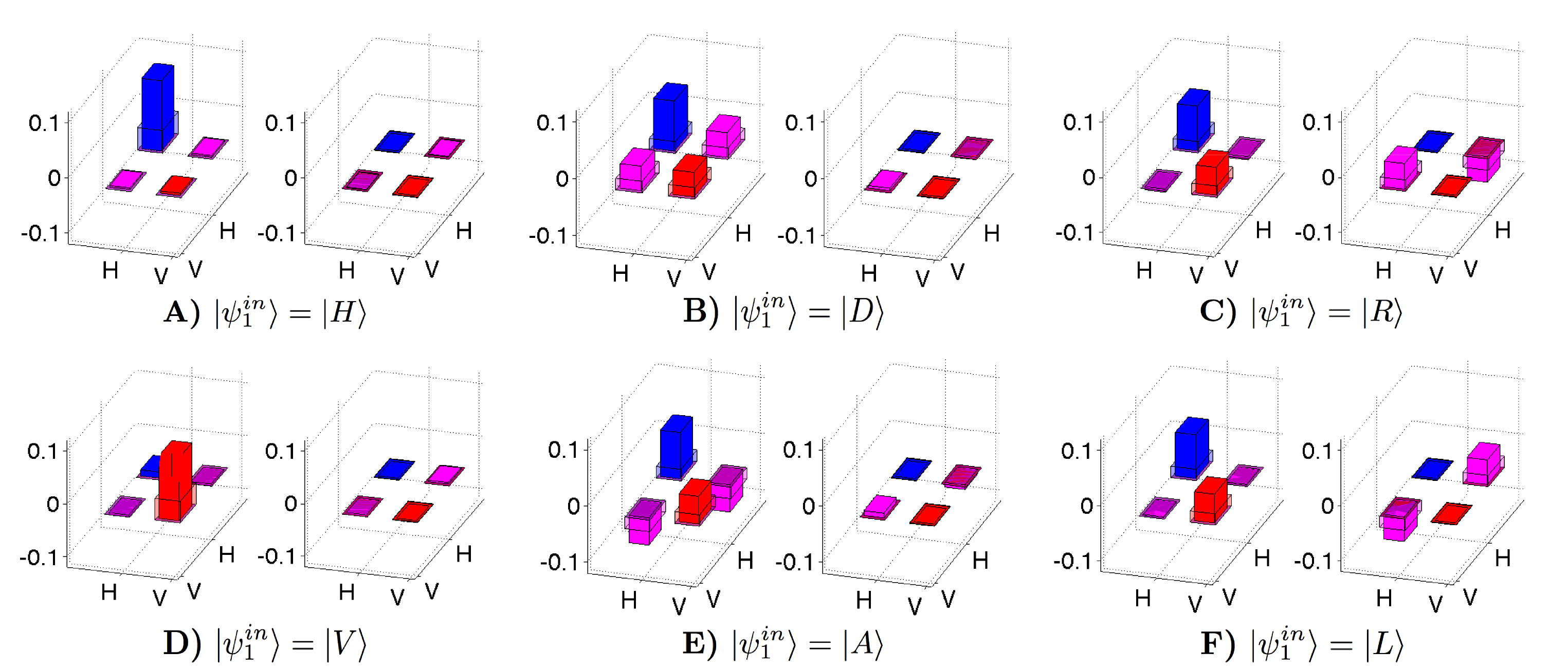}
\caption{Comparison of the qubit subspace for $\rho_{in}$ and $\rho_{out}$. Density matrix elements for the six canonical polarization inputs, with $G_m = 3.3 \pm 0.6$. The left graphs in subfigures A)---F) show the real elements of the density matrix, and the right graphs show the imaginary elements. The transparent bars are the state matrix elements of the single photon term $|\psi_{1}^{in}\rangle$ in the input state $\rho_{in}$, and the solid bars represent the amplified single photon term $|\psi_{1}^{out}\rangle$ of the output state $\rho_{out}$. The increase in the size of the single photon component in the mixed states is clearly apparent from the figures, as well as the fact that the coherences are preserved at the output of the circuit. A small systematic imbalance in favour of $|H\rangle$ is noticeable for all polarizations, and this is due to different heralding path efficiencies in the two NLA stages of the qubit amplifier.}
\end{figure*}

A great deal of attention has been devoted to the problem of efficiently transmitting quantum states--such as qubits--over significant distances. Some key examples serve to demonstrate why overcoming loss is of both fundamental and practical interest. From a fundamental standpoint, all long-range Bell inequality tests have been vulnerable to the \emph{detection loophole}: due to losses, not all entangled pairs are detected, and the \emph{fair sampling assumption} is invoked to argue that the undetected pairs would not have significantly changed the measurement statistics. Inevitable transmission losses can in principle be compensated by amplifying the signal. The theoretical limitation forbidding noiseless amplification of a quantum state can only be circumvented by making the process non-determinstic. Such a noiseless qubit amplifier, although non-deterministic, could amplify a quantum state in a \emph{heralded} way. A heralding signal allows two parties to be certain that they share a maximally entangled pair prior to measurement. This implies that the overall detection efficiency in the presence of heralding would no longer depend on transmission efficiency, but only on the intrinsic efficiencies of the measuring devices.  

Closing the detection loophole in an optical Bell test experiment is essentially equivalent to establishing device-independent quantum key distribution (DIQKD) between two parties, as the rigorous violation of a Bell inequality guarantees the presence of entanglement independent of the specific measurement procedure \cite{Gisin10,Lutken11}. Other approaches to overcoming the detection loophole have been proposed, such as heralding qubit states using quantum non-demolition (QND) measurements \cite{Kok02}, for example, but to date these other protocols have not been experimentally realised.

After transmission through any quantum channel with non-zero loss, a photonic qubit will be in the mixed state $\rho_{in}$, consisting of a vacuum and a single photon component,
\beq
\rho_{in} = \gamma_{0} |0 0\rangle \langle0 0| + \gamma_{1} |\psi_{1}^{in} \rangle \langle \psi_{1}^{in}| \ ,
\eeq
where the vacuum component will dominate ($\gamma_{0} > \gamma_{1}$) for a very lossy channel. The qubit is encoded in the polarization state of the single photon subspace:
\beq
|\psi_{1}^{in} \rangle = \alpha |1_{H} 0_{V} \rangle + \beta |0_{H} 1_{V} \rangle \equiv \alpha |H\rangle + \beta |V\rangle \ .
\eeq
The state $\rho_{in}$ is the input to the qubit amplifier. Such a heralded noiseless amplifier is a quantum circuit that works probabilistically, but with an independent heralding signal, and generates the transformation
\beq
\rho_{in} \rightarrow (1-P) |00\rangle \langle 00|\otimes \Pi_{f} + P \rho_{out} \otimes \Pi_{h} \ .
\eeq
Here $\Pi_{h}$ is the projector onto the subspace of heralding mode states corresponding to successful amplification, with the amplified state $\rho_{out}$ at the circuit output:
\beq
\rho_{out} = \frac{\gamma_{0} |0 0\rangle \langle0 0| + g^2 \gamma_{1} |\psi_{1}^{in} \rangle \langle \psi_{1}^{in}|}{N} \ ,
\label{eq:outstate}
\eeq
and $\Pi_{f}$ (fail) is the projector onto the subspace of cases where the heralding success signal is not received. The relative weighting of the qubit subspace $|\psi_{1}^{in} \rangle$ in the mixed state is increased by a factor $g^2$. In the absence of imperfections, the qubit amplifier leaves the qubit subspace itself unchanged; experimental imperfections may introduce some mixture to the qubit subspace so that the mixed qubit state $\rho^{qubit} \sim |\psi_{1}^{in} \rangle \langle \psi_{1}^{in}|$ replaces the perfectly pure $|\psi_{1}^{in} \rangle \langle \psi_{1}^{in}|$ in Eq. \ref{eq:outstate}. \blk Due to amplification, the output state must be renormalized by $N = \gamma_{0} + g^2 \gamma_{1}$.

\begin{table}[b]
\begin{tabular}{ccc}
$g^2$ \ \ \ & \ \ \ $G_{nom}$ \ \ \ & \ \ \ $G_{m}$ \\
\hline
\hline
$2.08 \pm 0.08$ \ \ \ & \ \ \ $2.0 \pm 0.2$ \ \ \ & \ \ \ $2.2 \pm 0.2$ \\
$3.48 \pm 0.09$ \ \ \ & \ \ \ $3.2 \pm 0.4$ \ \ \ & \ \ \ $3.3 \pm 0.6$ \\
$8.50 \pm 0.45$ \ \ \ & \ \ \ $6.5 \pm 0.8$ \ \ \ & \ \ \ $5.7 \pm 0.5$ \\
\hline
\end{tabular}
\caption{The nominal ($G_{nom}$) and measured ($G_{m}$) intensity gains were determined for three different splitting ratios $\eta_{H} = \eta_{V}$, with a qubit state size $\gamma_{1} = 0.041 \pm 0.005$. The nominal intensity gain was determined by measuring splitting ratios between the output detectors and heralding detectors in each NLA stage (see text and Appendix). The measured intensity gain was determined by taking the ratio of 
average detected photon number, conditional on heralding, at the circuit output, to the input state size $\gamma_1$.} 
\end{table}

With probability $P$, the transformation therefore increases the likelihood of detecting a single photon by a factor of $G_{nom} = g^2/N$, where $G_{nom}$ takes into account the renormalization. With probability $1-P$ the input state is transformed into the vacuum state, and the amplification fails. The maximum probability of success $P_{max}$ is bounded by the linearity of quantum mechanics. Due to the heralding, the case when amplification fails (the $1-P$ term in Eq. $3$) can be discarded, leaving only the state of interest $\rho_{out}$ to be sent on for further processing and measurement. Amplification occurs when $G_{nom} > 1$, implying that $\gamma_{0}/N < \gamma_{0}$, or that the vacuum component is reduced compared to that of the input state.

No experiments have previously been performed on the critical task of heralded qubit amplification. However, experiments on single-mode amplification have promised applications in continuous-variable entanglement distillation \cite{Xiang10}, continuous variable QKD \cite{Ferr10,Blandino12}, or enhancing the precision of phase estimation \cite{Usuga10}. A non-heralded experiment demonstrating the principle of loss mitigation for a single-rail qubit has also been reported \cite{Micuda12}. Nevertheless, for many quantum communication protocols, including BB84 \cite{BB84}, entangled-state protocols \cite{Gisin02}, and many other applications, the information will be encoded as a qubit in two optical modes. 

A heralded noiseless qubit amplifier may be constructed from two single-mode, noiseless linear amplification (NLA) stages \cite{Xiang10}, as observed theoretically by Gisin \emph{et al.} \cite{Gisin10}. These stages independently amplify the orthogonal polarizations $|H\rangle$ and $|V\rangle$ that are the basis states of the qubit. Although the stages are independent, their combined effect in the qubit amplifier is to establish coherence between two output modes that do not directly interact. 

The individual NLA stages are based on the quantum scissors \cite{Pegg98,Lvov03}, generalised such that the transmission of a central beamsplitter determines the nominal gain of the mode. The qubit amplifier circuit is shown schematically in Fig. 1. Successful amplification is heralded by detection of a photon by just one of the detectors in each of the two stages (either $D1$ or $D2$ in the first stage, and either $D3$ or $D4$ in the second stage), and the output state is analyzed using detectors $D5$ and $D6$. A key point is that the decision to keep particular signals is based solely on heralding events that occur before the final measurement basis choice. There is no post-selection based on the final measurement results.

Two pairs of single photons are generated from a type-I pulsed, double-passed, spontaneous parametric down-conversion (SPDC) source (see Appendix). The circuit employs three photons directly, leaving one photon as an external trigger. One of the single photons carries the qubit, and is sent through a highly reflective beamsplitter at the beginning of the circuit, to simulate a very lossy channel. The resulting mixed state becomes the input signal, $\rho_{in}$, to the amplifier. Two single photons are used as the ancillas, $|1_{H}\rangle$ and $|1_{V}\rangle$, that drive the NLA stages.

The loss in the signal mode was fixed, and the size of the single photon component in the mixed state, $\gamma_{1}$, was measured to be $0.041 \pm 0.005$ (see Appendix). The reflectivities of the central beamsplitters, $\eta_{H}$ and $\eta_{V}$, were calibrated by observing the ratio of detected single photons in $D6/D2$, for ancilla mode $|1_{V}\rangle$, and in $D6/D3$ for ancilla mode $|1_{H}\rangle$. The ratio $g^2 = \eta/(1-\eta)$ determines the nominal gain $G_{nom} = g^2/N$. In practice, successful amplification can be heralded by different combinations of detectors, and the observed splitting ratio $g^2$ (and hence the nominal gain $G_{nom}$) varied slightly with small differences in the path and detector efficiencies (see Appendix). The reflectivities $\eta_{H}$ and $\eta_{V}$ (Fig. 1) were set to be identical so that the gains of the two NLA stages would theoretically be equal.

\begin{figure}[t]
\includegraphics[width=\columnwidth]{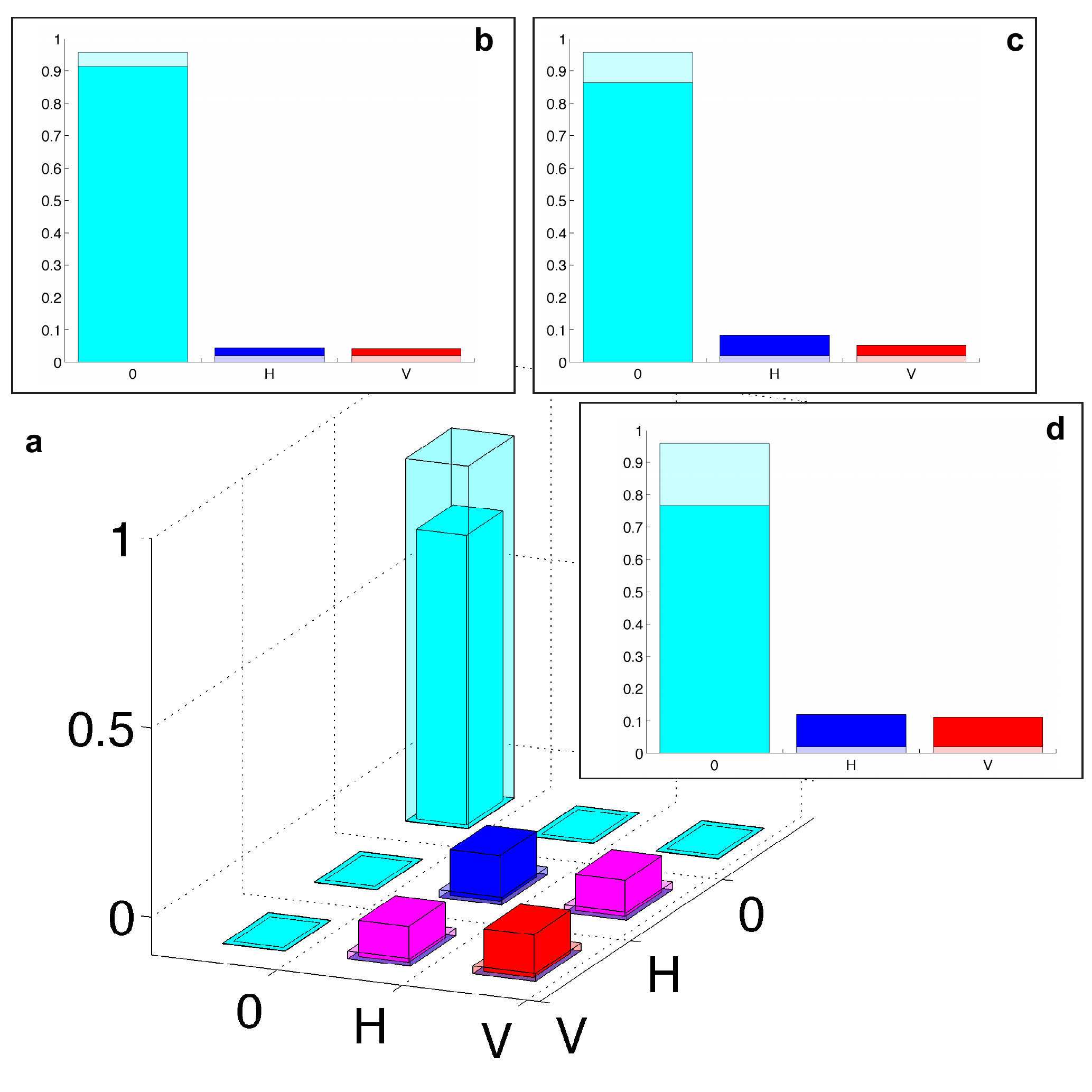}
\caption{{\bf{a)}} Comparison of the absolute value of the density matrix for $\rho_{in}$ and $\rho_{out}$.  The transparent bars are the absolute values of the matrix elements of the input state $\rho_{in}$, and the solid bars represent the amplified output state $\rho_{out}$, for gain $G_{m} = 5.7 \pm 0.5$ and input polarization $|\psi_{1}^{in}\rangle = |R\rangle$. 
{\bf{b)---d)}} Comparison of the $0$, $H$ and $V$ terms in the density matrices of $\rho_{in}$ and $\rho_{out}$, for the three different gains. The figure explicitly shows a decrease of the vacuum weighting in the mixed states, and a corresponding increase in the single photon intensities, as the gain increases.}
\end{figure}

\begin{table}[b]
\begin{tabular}{c|ccc}
$G_m$ \ & \ Input Fid. \ & \ Qubit Subspace Fid. \ & \ Output Fid. \\
\hline
\hline
$2.2 \pm 0.2$ \ & \ $0.041$ \ & \ $0.831 \pm 0.005$ \ & \ $0.072 \pm 0.001$ \\
$3.3 \pm 0.6$ \ & \ $0.041$ \ & \ $0.819 \pm 0.009$ \ & \ $0.119 \pm 0.008$ \\
$5.7 \pm 0.5$ \ & \ $0.041$ \ & \ $0.891 \pm 0.009$ \ & \ $0.208 \pm 0.002$ \\
\hline
\end{tabular}
\caption{For the case of all three measured gains, the fidelities between the amplified output states $\rho_{out}$ and the ideal qubit subspace $|\psi_{1}^{in}\rangle = |R\rangle$ were compared. The Output Fidelity is thus defined as Tr$[\langle R| \rho_{out}|R \rangle]$, the Input Fidelity is defined as Tr$[\langle R| \rho_{in} |R\rangle]$, and the Qubit Subspace Fidelity is defined as Tr$[\langle R| \psi_{1}^{out}\rangle \langle \psi_{1}^{out} |R\rangle]$.}
\end{table}

The input photon polarization state $|\psi_{1}^{in}\rangle$ was set to right-circular polarisation, $|R\rangle = \frac{1}{2}(|H\rangle - i|V\rangle)$, and the qubit amplifier was tested for three different nominal gains. The performance of the qubit amplifier was characterised in two ways: in terms of its measured gain, $G_{m}$, and in terms of the state fidelity between $|\psi_{1}^{in}\rangle$ and the amplified output state $\rho_{out}$. The measured intensity gain $G_{m}$ (Table 1) is defined as the ratio of average detected photon number after amplification to the average size of the qubit in the input state (see Appendix). A saturation effect can be seen when comparing the highest nominal gain setting $G_{nom} = 6.5 \pm 0.8$ and the corresponding measured gain $G_m = 5.7 \pm 0.5$---this is due to non-unit efficiency of delivering ancilla photons to the circuit \cite{Xiang10}.

For the case $G_{nom} = 3.2$, $|\psi_{1}^{in}\rangle$ was prepared in the six canonical polarization basis states \{$|H\rangle, |V\rangle, |D\rangle, |A\rangle, |R\rangle, |L\rangle$\}, and the density matrices of the output states were reconstructed using quantum state tomography \cite{White07}. The output state for the qubit subspace in each case is shown in Fig. 2. The fidelity between the output state and the input polarisation qubit, $\langle \psi_{1}^{in}| \rho_{out} | \psi_{1}^{in} \rangle$, was compared to the fidelity between the input state and the polarisation qubit, $\langle \psi_{1}^{in}| \rho_{in} | \psi_{1}^{in} \rangle$.  The fidelity, averaged over the six polarization states, increased from $4.1 \%$ to $11.7 \pm 0.8 \%$, for the measured gain $G_{m} = 3.3 \pm 0.6$. The increase in fidelity is slightly smaller than the value of $G_{m}$ would suggest, and this is because our amplifier introduces some polarisation mixture into the single-photon subspace $\rho^{qubit}$. This mixture is not a fundamental feature of amplification, nor is it due to source or detector inefficiency (see Appendix). Rather, it is a result of imperfect mode matching between the signal and ancilla modes, which translates to a decrease in the non-classical interference visibility, and hence an imprecise phase relationship. To a lesser extent, higher order photon terms from the SPDC source that populate the ancilla modes also contribute to polarisation mixture in the qubit subspace: one down converted photon in the ancilla pulse can trigger the heralding detector, and the other photon in the pulse can be directed to the output mode, without fixing the phase between it and the input mode. These unintended coincidence events look like polarisation mixture during state tomography of the output mode.

We compared the circuit output, with and without amplification, for the input state $|R\rangle$ in the case of the three gain settings. These data are shown in Fig. 3; the state fidelities between each of these three amplified states and $|R\rangle$ are presented in Table 2. The vacuum component of the output state is clearly reduced compared to the input state (Fig. 3), and there is a corresponding increase in the size of the single photon component. The purity of the polarization state remains high even after amplification (Fig. 2); there is a small variation in output state purities depending on the polarization input, and this was due to the fact that the two NLA stages had different HOM visibilities, and different efficiencies in the ancilla modes (see Appendix). 

This is the first experimental realisation of coherent amplification of a two-mode quantum state, which is an important advance towards meeting the open challenge of establishing DIQKD \cite{Gisin10}. The device achieves a significant improvement in transmission fidelity for qubits subjected to substantial loss, in a completely heralded way--no post-selection is employed. From theoretically investigating the effects of detection and source efficiency on the qubit amplifier's performance, we conclude that source inefficiency in the ancilla modes and lack of photon number resolution cause the gain saturation that we observe in our data (see Appendix). We show that in the $|g| \rightarrow \infty$ limit, the attainable gain in our circuit is in principle equal to the ancilla source efficiency. This is consistent with previous theory \cite{BerryLvov11}. Improved photon sources currently under development can be integrated directly with our device, and the circuit could therefore be used to amplify a state arbitrarily close to a single photon (\emph{i.e.} with arbitrary suppression of the vacuum), although amplification to this extreme level is not required to \textit{e.g.} violate a loophole-free Bell inequality. Heralded qubit amplifiers will have direct applications in DIQKD, fundamental tests of quantum physics, and a range of quantum technologies.

This research was conducted by the Australian Research Council Centre of Excellence for Quantum Computation and Communication Technology (Project number CE110001027). S. K. thanks D.~J. Saunders and M.~J.~W. Hall for useful discussions. 

\subsection{Appendix}
\section{\bf{SPDC source}} \\
We used a $2$mm thick $\beta$-Barium Borate (BBO) crystal, cut for type-I (polarization-unentangled) SPDC. The frequency-doubled output, at $390$ nm wavelength, of a mode-locked Ti:Sapphire laser was double-passed through the BBO crystal using a dichroic mirror, to generate two pairs of degenerate photons at $780$ nm. The pump power was kept constant at 100 mW, to limit the generation of higher-order photons in the ancilla modes of the circuit. 
\\ 
\section{\bf{State size and amplification measurements}}
\\
The signal and ancilla modes have either 0 or 1 photons per pulse, so determining the input signal size $\gamma_{1}$ and the amplified average photon number at the circuit output corresponds to determining the fraction of pulses, conditioned on the heralding signals of the two NLA stages, that contain a photon at the output. To measure the signal size, for example, the signal mode is transmitted directly through the circuit to detectors $D5$ or $D6$, without mode splitting or interference, and the ancillas are likewise transmitted directly to the heralding detectors ($|1_{V}\rangle$ to $D1$ or $D2$, and $|1_{H}\rangle$ to $D3$ or $D4$). The detected signal size is therefore the ratio of four-fold coincidences to three-fold coincidences: $\frac{C_{4}}{C_{3}}$, where $C_{3}$ is the appropriate combination of detected three-fold coincidences in ($D1$ or $D2$) $\&$ ($D3$ or $D4$) $\&$ the external trigger, and $C_{4}$ comprises an additional detection event in $D5$ or $D6$. To determine the state size at the amplifier \textit{input}, the detected state size is scaled by the detector efficiency ($\epsilon_{det}$) and path efficiency through the circuit ($\epsilon_{path}$). We use $\epsilon_{det} = 0.5$ for our avalanche photodiode detectors (Perkin Elmer SPCM-AQR-14FC) at $\lambda = 780$ nm, and we measured the average path efficiency from the circuit input to $D5$ and $D6$, $\epsilon_{path} = 0.64 \pm 0.04$. Therefore, the actual input state size $\gamma_{1} = \frac{C_{4}}{C_{3}} /\big (\epsilon_{det} \cdot \epsilon_{path})$.

The average amplified photon number at the output is measured using the same three-fold to four-fold ratio, but with the central beamsplitters in the NLA stages set to the correct reflectivities for amplification. Thus, $G_{m} = \big(\frac{C_{4}}{C_{3}}\big)^{amp}/\big(\frac{C_{4}}{C_{3}}\big)^{no \ amp}.$

All variable beamsplitters are implemented with a combination of half wave plates (HWP) and polarizing beamsplitters (PBS). The nominal gain $G_{nom}$ is measured for each NLA stage by comparing the ratio of detected singles in $D6$ or $D2$, for ancilla mode $|1_{V}\rangle$, and in $D6$ or $D3$ for ancilla mode $|1_{H}\rangle$. Since the detection efficiency varies for different paths through the circuit, the effective splitting ratios through all other paths in the circuit that could herald successful amplification were measured, to determine an average nominal gain for each NLA stage. In similar fashion, the variability in detection efficiency for different paths through the circuit was taken into account when determining the measured gain $G_{m}$, by measuring a representative sample of heralding combinations: detection in $D5$ or $D6$ heralded by $D1$ and $D3$, and detection in $D5$ or $D6$ heralded by $D2$ and $D4$. An average measured gain and a standard deviation were calculated using all the combinations.
\\
\section{\bf{Quantum state measurements}}\\
States within the qubit subspace of the output mode were determined using quantum state tomography. A small systematic single-qubit unitary operation imposed by the optical elements in our amplifier was corrected mathematically in producing the density matrices of Fig. 2 \& 3, and for calculating fidelities; in principle, this could be corrected using waveplates. The relative size of the vacuum component and the qubit subspace was determined from amplification measurements. Because the vacuum subspace arises from loss applied to a single photon, it is assumed that there is no coherence between the vacuum term and the single photon subspace.

The maximum attainable purity of a single--mode state at the output of a single NLA stage is limited by the HOM interference visibility at the central beamsplitter. The non-classical interference is measured in each NLA stage to characterise the mode matching between the signal and ancilla modes. In the first NLA stage, the interference visibility between the signal mode and $|1_{V}\rangle$ mode is typically $~99\%$---the signal and $|1_{V}\rangle$ are produced from the same pass of the double-passed SPDC source. In the second NLA stage, the interference visibility between the signal and $|1_{H}\rangle$ modes was typically $90-92\%$---the signal and $|1_{H}\rangle$ are produced in separate passes of the SPDC source, so this is an independent HOM interference \cite{Takeuchi12}. The maximum attainable purity of the polarisation qubit at the output of the amplifier circuit is therefore limited by the product of the two HOM interference visibilities.
\\
\section{\bf{Error analysis}}\\
Experimental uncertainties arise predominantly from two sources in our experiment: Poissonian counting statistics associated with the SPDC source; and averaging over variations in the path efficiencies for heralding with different detector combinations. This latter effect is primarily responsible for the error bars on the measured average gain values. Within the qubit subspace, however, the efficiency variation results in a decrease in fidelity due to slightly unbalanced amplification between $|H\rangle$ and $|V\rangle$ modes--that is, the measurement variation results in degraded performance rather than an uncertainty in the fidelity. The uncertainty in the fidelities of individual qubit subspaces is therefore dominated by Poissonian statistics. The error in the average state fidelity is dominated by the spread (which is nevertheless small) in the values for the six canonical polarizations.
\\
\section{\bf{Effect of imperfect detectors and sources on amplification}}\\
We consider the input state $\rho_{in}$ from Eq. (1), acted upon by a pair of identical NLA stages employing detectors with no photon number resolution and efficiency $\delta$, and single photon sources of efficiency $\tau$. The amplitude gain in the NLA stages is $g = \sqrt{\eta/(1-\eta)}$. A straightforward calculation shows that the un-normalized output state, for one of the four successful heralding signals, is given by the following expression:
\beq
\rho_{un} = \frac{\delta^2 \tau^2 (1-\eta)^2}{4} ((\gamma_{0} + L\gamma_{1})|00 \rangle \langle 00| + g^2 \gamma_{1}| \psi_{1}^{out} \rangle \langle \psi_{1}^{out} |) \ ,
\eeq
where
\beq
L = 1 + \frac{1-\tau}{\tau (1-\eta)} = \frac{1 + (1-\tau)g^2}{\tau} \ .
\eeq
The normalised output state is
\beq
\rho_{out} = \frac{(\gamma_{0} + L\gamma_{1})|00 \rangle \langle 00| + g^2 \gamma_{1}|\psi_{1}^{out} \rangle \langle \psi_{1}^{out}|}{\gamma_{0} + \gamma_{1}(g^2 + L)} \ .
\eeq
We briefly note a few features of the solution:
\\ (i) Detection inefficiency only reduces the probability of success of the qubit amplifier.
\\ (ii) Source inefficiency and lack of photon number resolution cause a gain saturation effect, denoted as $L$. They do not affect the purity of the qubit subspace.
\\ (iii) In principle, the best qubit efficiency that can be attained from the qubit amplifier is $\tau$, achieved in the $|g| \rightarrow \infty$ limit. We estimated that our average source efficiency, when factoring out the detector efficiency ($\epsilon_{det}$) and path efficiency through the circuit ($\epsilon_{path}$), to be approximately $0.45$.

The total probability of success is
\begin{equation}
P = \delta^2 \tau^2(1-\epsilon)^2(\gamma_{0} + \gamma_{1}(g^2 + L)).
\end{equation}
The experimental success probability was calculated from data by taking the ratio of three-folds heralding successful amplification, $C_{3}^{\ amp}$, and three-folds when the circuit is not set to amplify, $C_{3}^{\ no \ amp}$. This corresponds to the success probability conditional on an ancilla photon being delivered to the circuit and being detected, \textit{i.e.} with $\tau=\delta=1$. For the case $G_{m} = 3.3$, which is shown in Fig. 2, $P \approx 0.05$. This agrees with the expected value from Eq. 8.

\end{document}